\documentclass[12pt]{iopart}
\usepackage{color}
\usepackage{a4wide}
\usepackage{graphicx}
\usepackage{amssymb}

\begin{document}

\title{Distance dependence of the phase signal in eddy current microscopy}
\author{Tino Roll$^1$, Marion Meier$^1$, Ulrich Fischer$^2$, and Marika Schleberger$^{1*}$}
\address{$^1$Experimentelle Physik, Universit\"at Duisburg-Essen, D-47048 Duisburg, Germany}
\address{$^2$Physikalisches Institut, Universit\"at M\"unster, D-48149 M\"unster, Germany}
\ead{marika.schleberger@uni-due.de}

\begin{abstract}
Atomic force microscopy using a magnetic tip is a promising tool for investigating conductivity on the nano-scale. By the oscillating magnetic tip eddy currents are induced in the conducting parts of the sample which can be detected in the phase signal of the cantilever. However, the origin of the phase signal is still controversial because theoretical calculations using a monopole appoximation for taking the electromagnetic forces acting on the tip into account yield an effect which is too small by more than two orders of magnitude. In order to determine the origin of the signal we used especially prepared gold nano patterns embedded in a non-conducting polycarbonate matrix and measured the distance dependence of the phase signal. Our data clearly shows that the interacting forces are long ranged and therefore, are likely due to the electromagnetic interaction between the magnetic tip and the conducting parts of the surface. Due to the long range character of the interaction a change in conductivity of $\Delta\sigma=4,5\cdot10^{7}~(\Omega$m$)^{-1}$ can be detected far away from the surface without any interference from the topography.
\end{abstract}

\pacs{68.37.Ps, 68.37.Uv, 68.37.Rt}
\maketitle

\section{Introduction}
\label{intro}
Atomic force microscopy (AFM) \cite{binnig} has become one of the most versatile methods in surface science (see e.g. \cite{review}). In addition to the conventional topographic images it offers a huge range of information if the different acting forces can be exploited \cite{Gruetter, Hembacher, Sommerhalter}. In principle, any force acting on the cantilever may contain additional information about the sample. However, very often the question arises, what the relevant forces yielding the contrast in AFM images are and how they can be either accounted for and if or how they can be separated from each other. 

Obtaining information on the local conductivity of a sample can be achieved by using eddy current microscopy (ECM) \cite{Hoffmann1998,Hirsekorn1999,Lantz2001}. This method is based on the well-established method of dynamic scanning force microscopy (for a review see e.~g.~\cite{AFMbuch,Meyer2004}). In the latter, a cantilever oscillates above the sample thereby experiencing mainly two kinds of interaction forces. Apart from short-ranged chemical forces originating from the overlap of electron wave functions, there also exist long-ranged van der Waals-interactions arising from fluctuating induced dipole moments.
 
With ECM additional forces come into play. For example, when using a cantilever with a magnetic tip and a conducting sample, an additional dissipation channel occurs because the varying magnetization induces an eddy current in the sample giving rise to Joule heat dissipation. The same occurs for the inverse case of a conducting tip in combination with a magnetic sample - this time the eddy current is induced in the tip. If the microscope is operated in the so-called constant excitation mode, i.e. the excitation amplitude is kept constant during the scanning process, these energy dissipation channels manifest in a decrease in amplitude and an increase in phase-shift with increasing conductivity \cite{Hirsekorn1999}.

In this paper we present AFM(ECM)-data on the conductivity of especially prepared samples and demonstrate that the relevant force can be easily separated due to their long range character. 

\section{Experiment}
\label{sec:1}
\subsection{ECM with a conducting tip}
First experiments were performed under ambient conditions, using a commercially available atomic force microscope (AutoProbe CP-Research). As probes, typically Si cantilevers with resonant frequencies of $f_{0}=300$~kHz and a spring constant of $k=42$~N/m are used. To make the tips electrically conducting they are coated with a thin Pt/Ir film. The tip radius is about $r_{tip}=25$~nm. 

\begin{figure}[htb!]
\centering
\includegraphics[width=9cm]{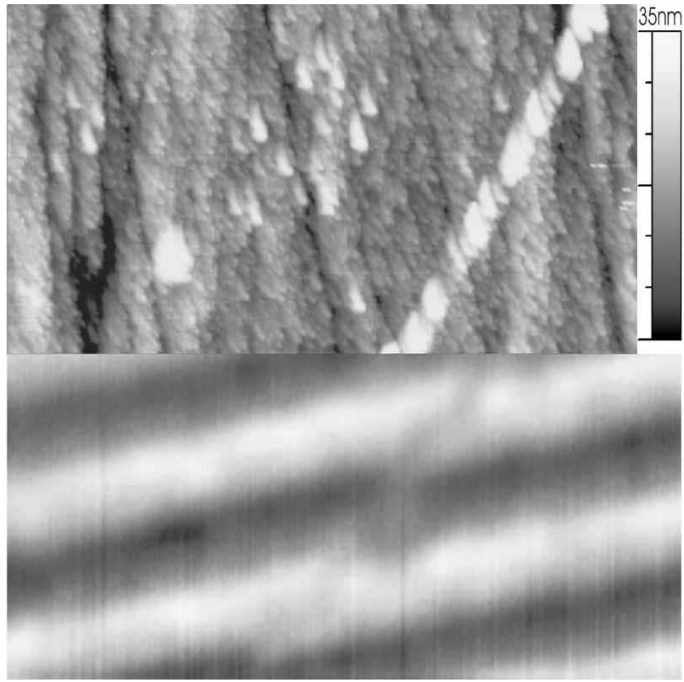}
\caption{Topography (top) and phase (bottom) image of a hard disc recorded with a conductive tip. Image size: 10 $\mu$m x 5 $\mu$m. Scan velocity $v_{s}=1$ Hz.}
\label{figure1}
\end{figure}

For the proof of prinicple we measured the surface of a commercial hard disc. In order to avoid a crosstalk of topographic features, it is important to sufficiently retract the tip before acquiring the phase signal. However, the microscope used for this experiment does not provide the possibility to manually define a specific distance between the tip and the surface. Therefore, the tip has been retracted from the surface until no contrast in the topography channel was visible anymore. Fig.~\ref{figure1} shows the obtained topography signal (top) at close distance as well as the phase image (bottom) at large distance, both measured with a non-magnetic (but conductive) cantilever. The straight scratch running from the bottom to the upper right corner is visible in both images. This remaining cross-talk shows that the tip is indeed still scanning the same spot on the surface (the sligt shift is due to the use of the coarse stepper motor to retract the tip for the phase image).

As can be seen, the phase image shows a clear contrast corresponding to the discrete magnetic tracks on the hard disc. The oscillating tip experiences a varying magnetic field originating from the stray field of the magnetic structures written onto the disc. This varying magnetic field induces eddy currents in the tip, which lead, according to Lenz's rule, to a damping of the oscillation amplitude and a shift in the phase between driving and resulting oscillation (see eq.~\ref{eq1}). Due to the fact, that the picture has been acquired in the amplitude modulation (AM) mode \cite{Lee2006}, the phase signal thus provides direct qualitative information on the local energy dissipation. Note, that contrary to a conventional MFM-measurement \cite{Gruetter1997}, the mapping of the magnetic structure shown here is done without using a magnetic probe.

\subsection{ECM with a magnetic tip}  
In order to study the conductivity of nanostructured surfaces, we focus on the inverse case of a magnetically coated cantilever oscillating in the near surface region. The eddy currents arise within the conducting sample and the fundamental physical principle is the same as described above. The AFM we used to perform these experiments was a Dimension 3100/NanoScope VI. With this microscope it is possible to perform more accurate measurements on the distance dependance of the phase shift induced by the electromagnetic interactions between tip and sample. The AFM was operated under ambient conditions in the AM mode. The cantilever (Veeco MESP-HM)has a magnetically coated tip (Cr/Co) with a radius less than $r_{tip}=10$ nm. The resonant frequency is $f_{0}=64$~kHz and the spring constant is about $k=3$~N/m. The magnetization is $M=400$~Oe which is equivalent to a magnetic moment of $m\geq3\times10^{-13}$~Am$^2$ (according to the manufacturer). The cantilever oscillates with an amplitude of $A=15$~nm. 

\begin{figure}[htb!]
\centering
\includegraphics[width=9cm] {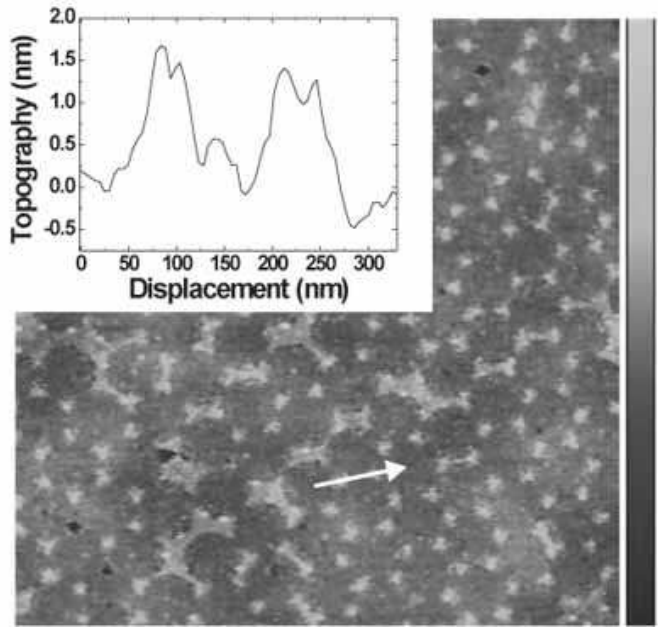}
\caption{Topography image of the gold pattern almost completely embedded in a polycarbonate matrix. Scan size: 2.5~$\mu$m x 2.5~$\mu$m, scan frequency: 1 Hz, Amplitude: 15~nm. The scale bar runs from 0--7~nm.}
\label{figure2}
\end{figure}

For this experiment, special samples are needed which are flat and at the same time exhibt a strong contrast in conductivity varying on a nanometer scale. 
Here, we use samples which were prepared as follows \cite{Fischer2002}. At first, a monolayer of hexagonally close-packed latex beads is formed on a polymer substrate. The Au projection pattern is produced by physical vapour deposition. Then, in a third step, the mask is removed without noticable remainders by floating the mask and the substrate onto a clean surface of water. Hence, we believe that patch charges play no role. This kind of sample is usually used as reference sample in scanning near field microscopy (SNOM). The triangular gold pattern left after the lift-off process provides us with a sample well suited for our experiment because it exhibits strong lateral variations in conductivity and has a relatively flat surface. The gold triangles with a conductivity of $\sigma=4,5\cdot10^{7}~(\Omega m)^{-1}$ are embedded in a non-conducting polycarbonate matrix and are only about 2~nm in height as can be seen from fig.~\ref{figure2}.

\begin{figure}[htb!]
\centering
\includegraphics[width=9cm]{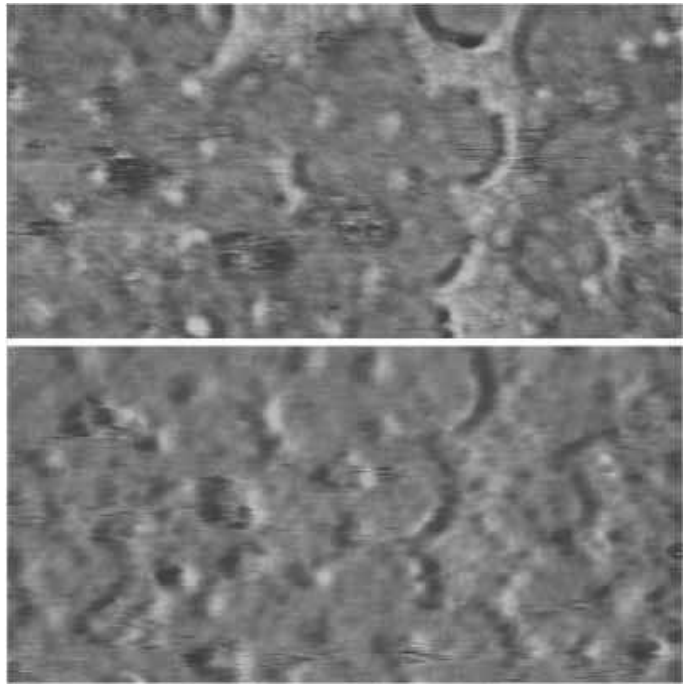}
\caption{Phase image of the gold pattern obtained at two different tip-sample distances of 40~nm (upper panel) and 100~nm (lower panel), respectively. The mean phase shift $\Delta{\phi}$ is $ 1.0^{\circ}$ and $ 0.3^{\circ}$, respectively. Image size: 1.5~$\mu$m x 0.75~$\mu$m. Note, that the scale in the upper panel covers $10^{\circ}$ and $5^{\circ}$ in the lower panel, respectively.}
\label{figure3}
\end{figure}

At first, we scanned the sample in the {\em tapping} mode, without increasing the tip-sample distance during the acquisition of the phase signal. A clear contrast between the polycarbonate and the gold areas is visible in the phase image (not shown here) as well as in the topography (see fig.\ref{figure2}). In order to determine the range of the forces acting on the cantilever due to eddy currents, we increased the tip-sample distance in steps of $\Delta{z}=10$~nm. For these measurements, the AFM was operated in the so called {\em lift mode}: during the first scan the topography is measured and stored. For the second scan the tip is lifted up to a predetermined height to exclude any effects from the topography. The lifted tip retraces the stored topography scan while the phase signal is recorded. In this way, long range forces can be separated from short-ranged ones. Fig.~\ref{figure3} shows typical phase signals obtained at two different tip-sample distances. Note, that the phase signal is still visible at a tip-sample separation of 100~nm.

\begin{figure}[htb!]
\centering
\includegraphics[width=11cm]{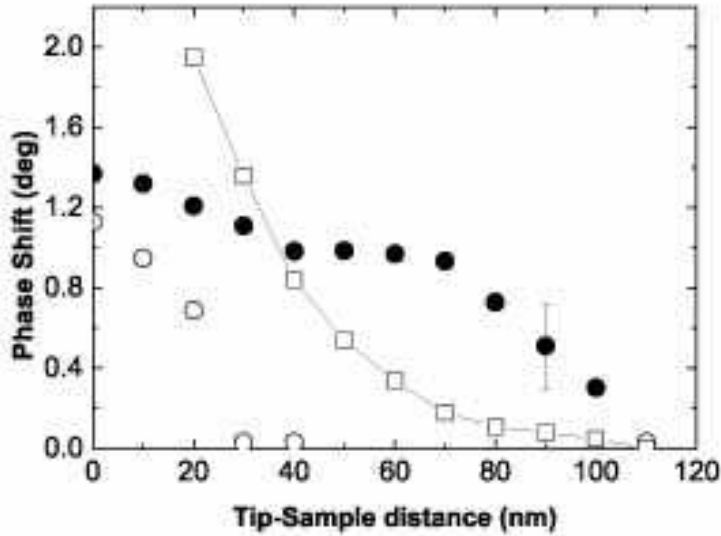}
\caption{Distance dependence of the phase signal for a magnetic (full circles) and a non-magnetic (open circles) tip. The error bar was determined from the {\it rms} values. The open squares denote the calculated phase shift using a dipole approximation (see eq \ref{eq1}). The line is drawn to guide the eye.}
\label{figure4}
\end{figure}
Fig.~\ref{figure4} shows the phase shift, determined from the {\it root mean square} ({\it rms}) roughness of the images, as a function of the tip-sample distance. The {\it rms} roughness provides a good measure of the mean phase shift for large tip-sample separations because it averages over all contributing features in the image and is not influenced by local variations of the surface (or the tip). The full circles in fig.~\ref{figure4} represent the phase signal acquired with a cantilever with a magnetic tip. For comparison, the open circles represent the phase signal measured with an uncoated, i.e non-magnetic tip (Nanosensors, NCHR, $f=300$~kHz, $k=42$~N/m). Both curves show a significant phase shift in the range of 1.0$^{\circ}$ to 1.3$^{\circ}$ for a tip-sample distance of about 10~nm. Note, that with an amplitude of $A=15$~nm we are still tapping the surface at that distance and crosstalk from the topography into the phase signal cannot be excluded. 

In the range from 10~nm up to 40~nm both the phase shifts for the magnetic as well as for the non-magnetic tip seem to decrease linear, though with a clearly different slope. At 30~nm, the phase shift in case of the non-magnetic tip is already zero, indicating that no long range forces are acting on the non-magnetic Si tip. At 40~nm the phase shift for the cantilever with a magnetic tip is still about 1.1$^{\circ}$ and seems to be constant up to a distance of 70~nm. At even larger distances the signal decreases linear again until it totally vanishes at a distance of about 110~nm. It becomes very clear from this diagram that long-range forces are acting on the magnetic tip. Because the gold triangles give rise to a clear contrast, we attribute these forces to the eddy currents induced in the conducting parts of the samples. Using the formula derived by Cleveland et al.~\cite{Anczykowsky1998}, we find that the phase shift corresponds to a dissipation power in the pW range.

\begin{figure}[htb!]
\centering
\includegraphics[width=9cm]{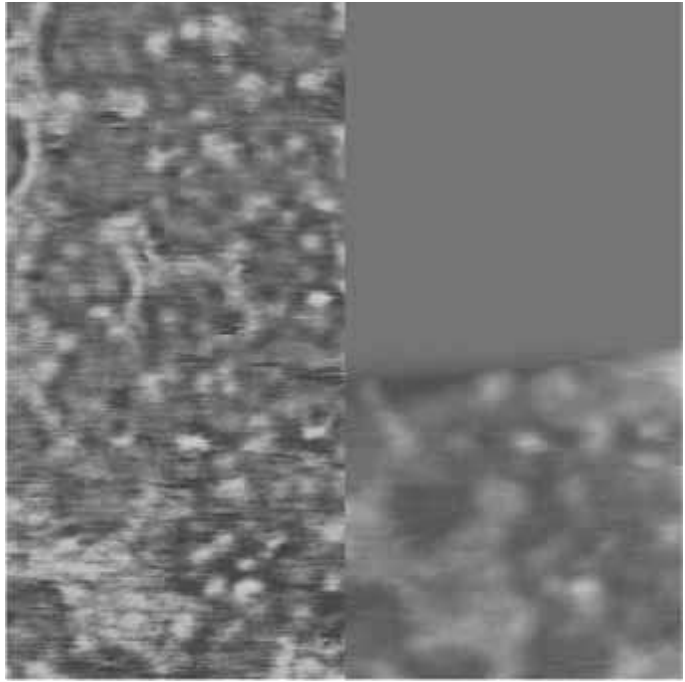}
\caption{Topography signal (right) acquired after tip was manually retracted until any signal disappeared and phase signal (left) which is recorded simultaneously. Image size: 0.75~$\mu$m x 1.5~$\mu$m. The phase scale in the left panel covers $10^{\circ}$. Scanning proceeds from top to bottom.}
\label{figure5}
\end{figure}

At large distances the amplitude signal is clearly different from the phase signal and may even disappear completely while the phase signal is still detectable. To demonstrate this we lifted the tip manually until the topographic image completely disappeared. The right panel of fig.~\ref{figure5} shows the acquired topography signal after the tip was retracted manually and the left panel shows the simultaneously measured phase signal. 

Comparing both images, one can see that in the upper area of the topographic image in fig.~\ref{figure5} there is no detectable signal. During the scan (running top-down) the tip-sample distance changes (either due to piezo-drift or more likely because the sample was tilted) and the topography begins to reappear. At the same time, the phase signal appears to remain nearly unaltered. Because of the large tip-sample separation both images appear rather diffuse. Nevertheless, we can determine the mean tip sample distance during the acquisition of the image in fig.~\ref{figure5} by using the data from fig.~\ref{figure4}. The mean separation in fig.~\ref{figure5} was thus between 40~nm and 70~nm in the upper part.

To describe the interaction between a magnetic tip and a conducting surface two different approaches have been proposed in literature. In the monopole approximation the sensor tip is long compared to the tip-sample distance, i.e. only one pole of an extended magnetic dipole is influencend by the stray field of the sample. The force is then given directly by the strength of the magnetic stray field. In case of the dipole approximation, the dimensions of the sensor tip are small compared to the distance between tip and sample. Here, the force is proportional to the gradient of the magnetic field. For a more detailed description of both approximations see e.g.\cite{Lohau1999}.

The phase shift we determined in our experiments is in good agreement with the results of the calculations from Hoffmann et.~al.~\cite{HoffmannDiploma,Hoffmann1998}. They estimated the effect using typical parameters for ambient conditions and found for a change in conductivity of $\Delta\sigma=10^{7}~(\Omega$m$)^{-1}$ a phase shift of approximately $1^{\circ}$. Calculations using a monopole approximation performed by Hirsekorn et.~al.~\cite{Hirsekorn1999} showed however, that typical cantilevers with a magnetic tip should in general not allow for the measurement of eddy current induced phase shifts because of the small magnetic fields emanating from such tips. In the same work though, the authors report that they observed a significant phase shift in Au/Ag samples and thus, the origin of contrast remained uncertain. 

Lantz et.~al.~demonstrated that it is possible to enhance the magnetic moment up to a moment of about $m=10^{-9}~$Am$^2$ by mounting a tiny FeNdBLa sphere on the tip \cite{Lantz2001}. This is however, very difficult to accomplish and the procedure seems not to be necessary when taking into account the good contrast already obtained with conventional tips.

If we use the same approach as Hoffmann we can estimate the phase shift in our experiment ($Q_c=20$, $A_0=15$~nm, $f=64$~kHz, $m\geq3\times10^{-13}$~Am$^2$, $\sigma=4,5\cdot10^{7}~(\Omega$m$)^{-1}$, the effective mass $m_{eff}$ of the rectangular cantilever which is about one-quarter of its real mass $m_c=4\cdot10^{-11}$~kg \cite{Rabe}, length = 225 $\mu$m, width = 25 $\mu$m, thickness = 3 $\mu$m) solving the differential equation for the cantilver oscillation at an average distance of $d_0=50$~nm:
\begin{equation}\label{eq1}	\frac{\partial^{2}d}{\partial{t}^{2}}+\frac{\omega_{c}}{Q_{c}}\frac{\partial{d}}{\partial{t}}+\omega^{2}_{c}(d-d_{0})+\frac{F_{z}(d,\partial{d}/\partial{t})}{m_{eff}}=A_{0}\omega^{2}_{c}cos(\omega{t)}
\end{equation}
with the Lorentz force
\begin{equation}	
{F_z}=-\frac{\sigma\mu^{2}m^{2}}{64\pi{d}^{3}}\frac{\partial{d}}{\partial{t}}\textbf{e}_{z}.
\end{equation}
This gives a phase shift due to eddy currents of about $0.5^{\circ}$ for a tip-sample distance of $d$=50~nm which is in agreement with the data shown in fig.~\ref{figure4} within a factor of 2. If we use the monopole approximation with our experimental parameters as input the phase shift is less than 0.01$^{\circ}$ and is thus in clear contradiction to our experimental data. The dipole approximation predicts a decreases with increasing distance over a range of $\approx$100~nm which is consistent with our data. The predicted slope follows the $1/d^{3}$ dependence of the Lorentz force while our data seems to decrease linear at distances larger than 60~nm. This indicates that terms might be present that are not taken into account correctly by the simple approximation used in the dipole approach. Thus the details of the contrast mechanism in eddy current microscopy need to be further investigated.

\section{Conclusions}
Atomic force microscopy using a cantilever with a magnetic tip and a nano patterned sample surface yields a significant phase shift for the conducting parts of the sample that is not connected to any topographic features. The value of the phase shift depends strongly on the tip-sample distance. We have shown that it is possible to measure a phase signal up to a tip-sample distance of about $100$~nm. Due to their long range nature we could identify the acting force being of electromagnetic origin. Therefore, by measuring the phase shift it is possible to distinguish between conducting and non-conducting areas of partially conducting samples at distances where the topography channel of the scanning force microscope exhibits no contrast anymore. 

However, phase images acquired in the AM mode are difficult to interpret. Therefore, in a forthcoming experiment we will investigate these SNOM samples under ultra high vacuum (UHV) conditions applying the force demodulation mode \cite{review} where the relation between the damping signal \cite{Anczykowski1999} and the eddy currents is more straightforward. This, in combination with the strongly increased $Q$ factor in UHV should enable us to study the characteristics of eddy current microscopy in a more detailed manner.

\section*{Acknowledgements}
Financial support by the DFG - SFB616: {\it Energy dissipation at
surfaces} is gratefully acknowledged.

\newpage
\section*{Figures}

\end{document}